# Injection induced seismicity size distribution dependent on shear stress

Yusuke Mukuhira*[1,2], Michael C. Fehler[1], Takatoshi Ito[2], Hiroshi Asanuma[3], Markus O. Häring[4]

[1]Earth Resources Laboratory, Department of Earth, Atmospheric and Planetary Sciences, Massachusetts Institute of Technology, Cambridge, Massachusetts, 02139, USA

[2]Institute of Fluid Science, Tohoku University, 2-1-1 Katahira, Aoba-ku, Sendai, 980-8577, Japan

[3]Fukushima Renewable Energy Institute, National Institute of Advanced Industrial Science and Technology (AIST), 2-2-9 Machiike-dai, Koriyama, Fukushima 963-0298, Japan.

[4]Häring Geo-Project, Wasserturmplatz 1, CH-4410, Liestal, Switzerland.

*Corresponding author: Yusuke Mukuhira (mukuhira@tohoku.ac.jp)

## Abstract

Like natural seismicity, induced seismicity caused by fluid injection also shows a power law size distribution, and its gradient *b*-value (ratio of small to large earthquakes) is often used for seismic hazard analysis. Despite well-known relationship that *b*-value is negatively correlated with differential stress for natural earthquakes, there is no understanding of the physical causes for *b*-value variations in injection-induced seismicity in the scale where the differential is nearly constant. We investigate a *b*-value dependence on the relative shear stress acting on existing fractures and show that the seismicity occurring along existing fractures with high shear stress have significantly lower *b*-values than does that associated with lower shear stress fractures. The *b*-value for injection induced seismicity is dependent on relative shear stress on faults. Our results provide a novel physical explanation for the *b*-value variations of induced seismicity.

1. **Introduction**

Earthquake size distribution follows a power law (Gutenberg and Richter, 1944) known as the Gutenberg-Richter (GR) relationship over a large range of earthquake scales from laboratory to megathrust, in addition to anthropogenic earthquakes. The gradient of the power law, the *b*-value or slope of that power law is obtained from the relation $\log_{10}(N)=a-bM$ (Gutenberg and Richter, 1944), where *N* is the number of events having magnitude at *M*, and *a* is a scaling factor. The low *b*-value means high relative number of larger earthquakes and vice versa. The *b*-value is often monitored for earthquake prediction and hazard analysis (Nanjo and Yoshida, 2018). Recent interest on this topic has led to several physical explanations for *b*-value variations (Ide et al., 2016; Nishikawa and Ide, 2014; Schorlemmer et al., 2005). Insights from various studies ranging from laboratory experiments (Amitrano, 2003; Goebel et al., 2013; Scholz, 1968) to natural earthquakes (Mori and Abercrombie, 1997; Spada et al., 2013; Wiemer and Wyss, 1997) have led to the assertion that *b*-value negatively correlates with differential stress (Scholz, 2015, 1968). Thus, spatial and temporal variations of *b*-value are generally understood to reflect spatial and temporal variations in the stress loading level.

Recently, the importance of induced seismicity to earthquake science and to society has increased along with the demand for subsurface development using fluid injection (Ellsworth, 2013; Evans et al., 2012; Majer et al., 2007). In the case of induced seismicity, *b*-value and its variation have also often been observed and used in various assessments of seismic risk (Bachmann et al., 2012, 2011; Langenbruch et al., 2018; Langenbruch and Zoback, 2016; McGarr, 2014; Shapiro et al., 2010). Though variations in *b*-value have often been observed in injection induced seismicity data (Bachmann et al., 2012), unlike natural seismicity, there is no general understanding of the physical causes for these variations and we have no basis for evaluating the importance of such variations (Gaucher et al., 2015). We cannot explain the *b*-value variations during induced seismicity as being caused by variations in differential stress as we do for natural earthquakes because the time scale, tectonic setting and the stress state of the fractures are different from natural earthquakes. The region activated by fluid injection is on the scale of several km and its depth rarely exceeds 5 km. A significant variation in differential stress is not expected on these scales. The static stress change caused by earlier induced seismicity can change stress state locally, but in the induced seismicity case, these changes are small so their effect on induced seismicity is much less significant than pore pressure (Catalli et al., 2013), which does not alter differential stress. Therefore, the question we address is what is the physical explanation of *b*-value variation for injection induced seismicity?

During hydraulic stimulation, shear slip occurs along fractures having a range of orientations. The geometrical relationship between an existing fault and *in-situ* stress determine the stress state of the fault, which is the balance of shear stress and normal stress. Therefore, there is some variation of shear stress acting on the existing faults which are candidate locations for induced seismicity according to the geometrical relationship to the *in-situ* stress and fault orientation (see methods). The required pore pressure increases to destabilize existing fault varies by its stress state.

In many *b*-value studies in natural earthquake settings (Mori and Abercrombie, 1997; Nanjo and Yoshida, 2018; Scholz, 2015; Schorlemmer et al., 2005) and laboratory experiments (Amitrano, 2003; Goebel et al., 2013; Scholz, 1968), *b*-value variation has been found to correlate with the strength, differential stress, and shear stress. First, we might consider the correlation to the stress relative to shear strength (Scholz, 1968) in the form of Coulomb failure stress (CFS) or pore pressure increase. However, if we assumed that all shear failure is caused by pore pressure and that the static stress change is negligible condition, CFS changes only due to the pore pressure increase in injection-induced setting. Pore pressure promotes shear slip by reducing the effective normal stress but it does not affect shear stress. In other words, pore pressure moves the Mohr stress circle closer to the failure line, but pore pressure does not change the differential stress, which is the diameter Mohr stress circle. Therefore, for induced seismicity studies, we cannot consider CFS or pore pressure change as a substitute for differential stress change as used in natural earthquake studies.

Instead of using differential stress or CFS for a natural earthquake case, we investigate the relationship between *b*-value and the shear stress on existing faults that slip. We consider that the variation of shear stress acts as driving potential controlling the size of the slip. We use an integrated analysis of microseismicity data and measured *in-situ* stress information to infer the shear stress along the existing faults that slip. Also, this is the first case study where the *b*-value dependency on the stress is confirmed for field seismicity and directly measured *in-situ* stress data beyond the laboratory scale.

## 2. Study field and Data

### 2.1 Study field: Basel, Swizterland EGS project

We use the induced seismicity data recorded during the Basel Enhanced Geothermal System (Häring et al., 2008) hydraulic stimulation since high-quality microseismic data and *in-situ* stress information are available. Hydraulic stimulation was performed in the Basel urban area, in December 2006. Basel is located near the border of Switzerland, Germany and France, where is the southern end of the Rhine Graben. This area is thought to be the area of highest geothermal potential in northern Europe (Häring et al., 2008); where other classic EGS projects were operated (Evans et al., 2005). Injected water penetrated into the granite basement throughout the permeable zone in an open-hole section (4603-5000 m) of the injection well; Basel-1 (true vertical depth; TVD 5000 m) (Häring et al., 2008). The flow rate was increased step by step and finally reached to 3300 L/min on the 5$^{th}$ day of the stimulation at which time the wellhead pressure (WHP) was 29.6 MPa. Hydraulic stimulation successfully reactivated a number of existing fractures causing microseismicity from the start of injection. The locations of the microseismicity indicated that an artificial geothermal reservoir was created successfully by enhancing the permeability in the vicinity of the injection well.

**2.2 Microseismic data**

The operating company deployed a microseismic monitoring network that consisted of six borehole stations and one temporary station in the injection well. All stations had 3-component seismometers. Microseismic monitoring station locations are shown in Fig. 1(d). Using the waveforms from 6 downhole stations, we determined the hypocenters of 2800 microseismic events from manual picks having high confidence (Asanuma et al., 2008). Our target events occurred from the start of the injection until after 6 months from the start of the injection (Häring et al., 2008; Mukuhira et al., 2013). We then performed a cluster analysis based on waveform similarity (Moriya et al., 2003) and consequently relocated hypocenters of 60 % of the microseismic events (Asanuma et al., 2008) with the double difference method (Waldhauser and Ellsworth, 2000) (Fig. 1). The seismically activated region is roughly a cube 1,000 m on a side and the reservoir depth is 4,000~5,000 m in the granite basement (Fig. 1).

**2.3 *In-situ* stress data**

The orientation of $S_{H\max}$ (maximum horizontal stress) has been well constrained by borehole breakout analysis for well Basel-1. The mean orientation of $S_{H\max}$ (maximum horizontal stress) in the basement granite is N144°E±14° (Valley and Evans, 2015, 2009). Stress magnitude was also estimated from borehole breakout analysis using Schlumberger ultrasonic borehole televiewer (UBI) logging run data taken from the granite section (2569-4992 m) of Basel-1 (Valley and Evans, 2015). The consistency of the stress information results was investigated by comparing with other geophysical observations. The consensus estimates of in-situ stress magnitude as a function of depth z in meters from the free surface are (Valley and Evans, 2015):

$$S_{Hmax} = 0.00104z + 115 \tag{1}$$

$$S_{hmin} = 0.01990z - 17.78 \qquad (2)$$

$$S_v = 0.0249z \qquad (3)$$

where stress is in MPa. The stress magnitudes given by equations (1-3) are shown in Fig. 2(a). $S_{Hmax}$ has a very low variation with depth, and this causes a transition from a strike slip to a normal faulting stress state regime at around 4800 m depth. Six depth dependent Mohr stress circles at 200 m intervals in the 4,000-5,000 m depth range are presented in Fig. 2(b). The Mohr stress circles become smaller with depth meaning that differential stress decreases with depth. However, it varies from 60 MPa at 4000 m to 40 MPa at 5000m, which is quite small and not sufficient to cause a significant change in differential stress variation since the differential stress discussed in natural earthquake is much larger than this value (Scholz, 2015).

**2.4 Fault orientation data**

Orientations of fault planes are obtained from the focal mechanisms estimated for about 100 of the relatively large events using data from the Swiss natural seismicity network (Deichmann and Giardini, 2009; Terakawa et al., 2012). We select the slip plane using geomechanical information and confirmed them with the hypocentral distribution of other microseismicity in the vicinity of those lager events (Mukuhira et al., 2017). The poles to the selected fault planes are shown in Fig. 3(a).

To add more fault orientation information, we then extracted the fault orientation information from the result of the seismic cluster analysis. Clustering analysis of microseismic events is based on waveform similarity obtained from coherence. We expect that the hypocenters of events having similar waveforms are located close to each other and that they have the same source mechanism (Moriya et al., 2003; Shelly et al., 2016). The locations of events in many clusters delineated a planer or streak shape that are visualized in Fig. 1. Then assuming that all events from each cluster have the same orientation, we applied principal component analysis (PCA) to the hypocenter distribution of each of the clusters. We estimate the orientation of fault from the normal vectors (3rd eigen vector) of 100 seismic cluster faults. Fig. 3(b) shows the normal vectors of three eigen vector resulted from PCA analysis. Fig. 3(c) shows the frequency distributions of the strikes of events in seismic clusters (purple) and Fig. 3(d) shows the strikes of seismic clusters themselves (orange). A planer or streak shape of the seismic cluster are again confirmed by contribution ration of eigen values shown in Fig. 3(e). To exclude unreliable information due to multiplets that have small numbers of events, we used only those that consisted of more than 5 events.

In comparison of poles distribution from FPS and cluster analysis (Fig. 3(a) and (b)), the green dots showing the normal vectors of luster situate in the region where the poles of FPS exist, showing over all consistency between fault orientations from FPS and cluster analysis. In addition to the pole plot, the rose diagrams in Fig. 3 (c) and (d) show that the strikes of the clusters are estimated quite reasonably as they distributed within 0~30 degrees from the orientation of $S_{Hmax}$. Consequently, around 1000 out of 2800 events are available for shear stress analysis.

## 3. Methods

3.1 Geomechanical analysis

Using the *in-situ* stress model and orientation of an existing fault, we estimate the shear stress on the fault; shear stress $\tau$ and normal stress $\sigma_n$ working on a fault plane with a known orientation can be determined from the magnitude of the principal stress $S_n$ ($n$=1, 2, 3; $S_1 > S_2 > S_3$) and their orientations, and geometry of the fault plane as:

$$\tau = \{a_1^2 a_2^2 (S_1 - S_2)^2 + a_2^2 a_3^2 (S_2 - S_3)^2 + a_3^2 a_1^2 (S_3 - S_1)^2\}^{\frac{1}{2}}, \tag{4}$$

$$\sigma_n = a_1^2 S_1 + a_2^2 S_2 + a_3^2 S_3, \tag{5}$$

where $a_n$ ($n$=1, … ,3) are direction cosines from each principal stress to the vector normal to the fault plane.

Due to the state of stress transitions and the differential stress perturbation with depth,

the absolute value of shear stress can be influenced by the differential stress and state of stress. To remove those factors, we introduce a measure called Normalized Shear Stress (NSS) instead of shear stress, which is the ratio of the shear stress on a fault to the maximum shear stress at the fault's depth. Stress state for the fault with maximum shear stress is the top of the Mohr stress circle, so the maximum shear stress is expressed as a radius of Mohr stress circle. So, NSS corresponds to the ratio of the height of the point showing the stress state of the fault on the Mohr stress circle to the radius of the Mohr stress circle.

**3.2 *b*-value analysis**

We employ standard and well accepted method to estimate *b*-value. In following section, we divide the catalog by NSS value and then estimate Mc for each sub-catalog using the entire magnitude range method (EMR) (Woessner and Wiemer, 2005). We estimate *b*-value with maximum likelihood method (Aki, 1965; Utsu, 1999) and binning size of magnitude in *b*-value estimation is 0.1. All computations were performed using ZMAP (Wiemer, 2011). To test the discrepancy of the b-values in sub-catalogs divided by NSS, we employed the Utsu's test (Utsu, 1999). The equation is in Appendix.

**4. Results**

**4.1 Correlation between NSS and magnitude**

First, we simply examine the correlation between the magnitude of seismic events and NSS (Fig. 4). We observe a correlation between the magnitude and NSS with the large events showing a tendency to occur along the faults with higher NSS. Smaller events also occur along faults with high NSS, so the correlation between event size and NSS is not a simple one. In order to investigate the relation in a statistical sense, we introduce the concepts of *b*-value and investigate the relationship between *b*-value and shear stress.

We divide the catalog by NSS threshold value into those with NSS higher than the threshold as the higher group: HG and those with lower NSS as the lower group: LG (see Fig. 4). We plot the GR magnitude frequency distribution of the events and estimate the Mc with the EMR method (Woessner and Wiemer, 2005), and the *b*-value with the maximum likelihood method (Aki, 1965; Utsu, 1999). Then *b*-values for both groups are compared in Fig. 5 (b) and (e). We show the power law slopes defined with estimates of *b*-value and constant *a* in comparison with binned numbers of events (Main, 2000). The binned number of events (triangles) for the HG scatters, especially on the sub-catalog for the HG (0.87), where the total number of events in sub-catalog is smaller than the other HG data sets. We generated 1000 synthetic catalogs with a single power law having our estimated *b*-values, and we show their binned frequency distributions with yellow markers in Fig. 3 (Naylor et al., 2009). With the exception of very few samples in the HG data, the binned counts of our real data fall within the range of statistically possible variation. Visual fit between power law slopes to the binned number of events for LG data are always good (Fig. 5 (d)-(f)) and all plots of binned number are within the range of statistical variation reasonably. These observations confirm that the sub-catalogs created by dividing data by NSS thresholds still follow a single GR power law.

We first look the *b*-value dependency of HG using NSS thresholds of 0.71, 0.76 and 0.87, which correspond to the NSS of well oriented fractures having friction coefficients of 0.6, 0.85, and 1.0 respectively (Fig. 5(a)-(c)). We observe that *b*-value for NSS > 0.71 is higher than that for NSS > 0.76 and 0.87, but *b*-value for NSS > 0.76 is lower than that for NSS > 0.87. Meanwhile, for the LGs (Fig. 3d-f), *b*-value systematically decreases with increasing NSS threshold from 1.31 for NSS < 0.71 to 1.12 for NSS < 0.87. The other observation made from the comparison of *b*-values for the same NSS threshold (e.g. Fig. 5(a) and (d)) is that the HGs show significantly lower *b*-values than those of LGs.

## 4.2  *b*-value dependency to normalized shear stress

We further investigate the relationship observed in Fig. 5, by determining *b*-value for a range of NSS thresholds. We estimate Mc for each sub-catalog defined by NSS threshold and estimate corresponding *b*-values (Fig. 6). Overall dependency of *b*-values on NSS threshold is clearly observed for every 0.05 increment in the threshold for HGs (Fig. 6(a)) and LGs (Fig. 6(b)). We estimate *b*-values when a sub-catalog contains more than 100 events ≥ Mc. The overall trend of *b*-value is to decline with increasing NSS threshold. As we estimate Mc for each sub-catalog according to NSS, so Mc also varies as we can see in Fig. 6(c), which possibly leads to a slight fluctuation of *b*-value trend. The *b*-value for all available events is 1.07. The *b*-values for HG events are always less than this value. On the other hand, the *b*-value of LG groups more systematically decrease with increasing NSS threshold and finally converge to the *b*-value for all available events when the threshold is 0.95. Due to small number of events, we cannot estimate Mc nor *b*-value for NSS < 0.55. The *b*-values from LGs become stable as the threshold increases because of the increase in the number of events used for the estimation (Fig. 6(e)). Consequently, both cases show a *b*-value dependency on NSS threshold. Meanwhile, we observe that the *b*-values for HG groups are significantly lower than those for LG groups with the same threshold. The significance of the differences between them is evaluated with statistical test (Utsu, 1999) and they show (Fig. 6(d)) that differences are highly significant (AIC = 5 is highly significant line).

### 4.3 Uncertainty

We discovered the b-value dependency on NSS based on the b-value and stress analysis using the in-situ stress model, which we assume laterally constant stress condition. Although we have evidence for laterally constant stress model such as the consistent orientation of SHmax from the injection well and adjacent monitoring well (Valley and Evans, 2009), and systematic pattern of microseismic occurrence and their distribution (Mukuhira et al., 2013),we examine the effect of uncertainty in stress model to the b-value analysis. We synthesize 1000 realization of stress models which contain 10 % error (2s) following a normal distribution in SHmax and SHmin and normally distributed error (2s=±14°) in the orientation of SHmax. We compute b-values for each HG and LG to various thresholds with constant Mc (0.8) for the stability of analysis. Fig. 7 shows 500 examples of b-value estimates from the stress model considering uncertainty and the average b-value from 1000 trials. Note that errorbars in Fig 7 show 2 standard deviations (2s) estimated from the distribution of b-value estimates, and they don't show the uncertainty in b-value estimation.

We can observe that both b-values for HG and LG decrease with NSS, though the b-value for HG slightly decreases. In addition, b-values for HG are lower than b-value for LG except for the points more than 0.9, where we cannot estimate b-value for HG due to a small number of the events. We can confirm three main observations we had for the observed stress model, even when we employ the stress model considering the stress heterogeneity in magnitudes and orientation.

We further examined *b*-value dependency on shear stress with Mc estimated with another method (maximum curvature method), constant Mc, and intentionally overestimated Mc and confirmed our findings (Supplementary Fig. 1 - 5). We also performed a statistical simulation of the influence of the approximately 1400 events for which we cannot determine NSS. We found that our observations and conclusions are not changed (Supplementary Fig. 6 – 8).

5. **Discussion**

Our main observations are that grouping events using both upper and lower bounds on NSS show *b*-value variation associated with changes in NSS threshold and that the HG groups always show lower *b*-values than those estimated for LG groups with the same NSS value. Our observations demonstrate a clear *b*-value dependency on NSS for injection induced seismicity. The discovered *b*-value dependency seems to be linked to the general interpretation of b-value dependency on differential stress, in terms of the dependency on the stress. However, the interpretation of our discovered dependency is slightly different from the general interpretation of *b*-value dependency.

We first consider the general interpretation of *b*-value dependency on differential stress using the Mohr stress circle. The Mohr stress circle shown in Fig. 8(a) shows how the diameter of the circle changes with increasing differential stress. The *b*-value negatively correlates with differential stress, i.e., the diameter of the Mohr stress circle. Many studies of the *b*-values of natural earthquakes found indirect indicators of differential stress such as depth (Mori and Abercrombie, 1997), stress regime type (Schorlemmer et al., 2005), and subduction plate age (Nishikawa and Ide, 2014). Other *b*-value studies from laboratory experiments found a *b*-value reduction with increasing loading stress (Goebel et al., 2013; Scholz, 1968). These studies implicitly or explicitly consider that most of the earthquakes occurred along the representative fault or fractures that are well oriented for slip under the given stress condition. Thus, we indicate the stress state of well oriented faults with dots on the Mohr stress circle in Fig. 8(a). As differential stress increases, the shear stress along well oriented faults increase as well. So, the general interpretation of *b*-value dependency on differential stress can be equivalent to the *b*-value dependency on the shear stress (here absolute shear stress). Some of the *b*-value studies discussed *b*-value dependency with shear stress (Ide et al., 2016; Nishikawa and Ide, 2014; Yoshida et al., 2017).

Let us recall our geomechanical setting considered in this study. On the reservoir scale, differential stress does not change significantly with time or space. Therefore, the differential stress is mostly constant, leading to a constant diameter of the Mohr stress circle (Fig. 8(b)). In the reservoir, various type of the existing fault is there and stress state of them also varies according to the geometrical relationship between faults and *in-situ* stress. The existing faults can have shear slip according to the pore pressure increase by fluid injection to destabilize it. During the stimulation, Coulomb failure criterion line moves to the right in relation to the Mohr stress circle due to the injection pressure increases (this is equivalent to Mohr stress circle shifting to the left in the case where the horizontal axis is effective normal stress). As this figure shows, pore pressure increase does not affect to differential stress nor shear stress. Pore pressure increase just prompt the failure but the potential of shear slip (differential stress) remains as before. Then, the faults on the left of the failure line reach a critical state and fail. Here, we can see the difference in the process to reach the failure. In natural earthquake case, differential stress increase leads the failure because of increasing loading, so Mohr stress circle getting bigger. Meanwhile, induced seismicity case, pore pressure increase leads the failure because of pore pressure increase which decreases effective stress, so Mohr stress circle shift to the failure line.

As we explained NSS is the relative height of Mohr stress circle for the point that shows the stress state of the fault. Then, stress states of the high NSS faults are situated in the higher portion and critical part of the Mohr stress circle (black shaded) and these are the events that lead a lower *b*-value (Fig. 8(b)). Thus, when more events occur within the higher shear stress fault, the *b*-value decreases. Conceptually, this relationship is shown by the upward pointing yellow arrow in Fig. 8(b) suggesting that the more events from high shear stress fault cause the lower *b*-value. This is consistent with the conceptual explanation of *b*-value dependency on applied shear stress with upward yellow arrow shown in Fig. 8(a) suggesting the (absolute) high shear stress cause the lower *b*-value. So, we could explain the ubiquitous phenomena of *b*-value variation but in the case of injection induced seismicity, where tectonic stress condition and triggering process are totally different, with a different theory compared to the general understanding in seismology. At the end, our explanations are quite consistent with the insight of *b*-value variation. Also, the larger event occurring from relatively higher shear stress fault is reasonable since the higher shear stress fault has much potential to the size of shear slip.

One might consider that pore pressure increase correlates to the b-value variation as the frictional strength does (Scholz, 1968). CFS is also used as the stress relative to the failure in natural earthquake case. Pore pressure and CFS are essentially equivalent in the injection induced seismicity setting since shear and normal stresses are constant, and pore pressure is the only variable parameter in the definition of CFS (Beeler et al., 2000). As we discussed, the pore pressure and CFS cannot change the differential stress and shear stress. So, b-value variation interpretation with pore pressure is not consistent with general interpretation. This hypothesis should be investigated with data analysis in future.

The largest event in this field and several other large felt events occurred in the deeper part of the reservoir where the differential stress is slightly smaller than at shallower depths and where the state of stress is normal fault slip type. From previous studies, the larger events would thus be expected to have occurred within a strike slip stress regime rather than the normal fault slip stress regime (Schorlemmer et al., 2005). Our results show opposite tendency suggesting that the differential stress interpretation behind of the b-value dependency on stress regime is not the case. In addition, we do not observe a significant preference for larger events to occur shallower in the reservoir where differential stress is larger (Mukuhira et al., 2017). From these observations, there is a tendency for the occurrence of large events to be correlated with normalized shear stress rather than differential stress.

## 6. Conclusions

We conclude that, on the reservoir scale, a lower *b*-value associated with fluid injection can be attributed to the occurrence of events along fractures that are oriented to have high shear stress. Our conclusion provides the first clear physical interpretation of variation in *b*-values for induced seismicity where earthquakes occur along fractures of various orientations in a constant regional stress condition. Also, this is the first case to show the *b*-value dependency on the directly measured stress data beyond the laboratory scale. This helps to fill the gap in our understanding of *b*-value between the laboratory scale and earthquakes along plate boundaries or in the continental crust, clarifying that even for the microseismicity that occurs within one local region, *b*-value can be related to shear stress level.

Our findings contribute to our ability to assess seismic risk and to mitigate it when undertaking subsurface development associated with fluid injection since the *b*-value reduction can to some degree be controlled by the injection pressure. With an increase of injection pressure, the Coulomb failure criterion line shifts to the right (Fig. 5(b)), meaning that the range of orientations of fractures that can have shear slip increases. A higher injection pressure generally increases the chance of slip along high shear stress fractures, which, due to the lower *b*-values of these events, are likely to have larger magnitude. Note that reactivation of an existing fault is determined by the balance between shear and normal stress. Thus, a well-oriented fault is not the identical to one having maximum shear stress. A high shear stress fault would fail after sufficient increase of pore pressure according to the normal stress on that fault. In fact, it has been reported that many of the large events accompanying fluid injections have occurred when injection pressure reached its highest level as in the Basel field (Mukuhira et al., 2013). Though there are other factors that control the size of the largest injection induced earthquakes, our findings provide the basis for understanding of large induced seismicity and how to control the occurrence of large induced seismicity: avoiding high injection pressure that can induce events along faults with higher shear stress.

# Appendix

**Significance test of difference in *b*-values from sub-catalogs (Utsu's test)**

For the two groups of earthquakes having $N_1$ and $N_2$ events and *b*-values $b_1$ and $b_2$ respectively, the significance of the difference in *b*-value is tested with Akaike's Information Criterion (AIC). ΔAIC is given by

$$\Delta AIC = -2(N_1 + N_2)ln(N_1 + N_2) + 2N_1 ln\left(N_1 + N_2\frac{b_1}{b_2}\right) + 2N_2 ln\left(N_1\frac{b_2}{b_1} + N_2\right) - 2$$

(A1)

ΔAIC=5 means that the difference in *b*-values between the two data sets is highly significant and ΔAIC=2 means the difference is significant(Aki, 1965).

**Acknowledgement**

We thank R. Abercrombie, C. Dinske, B. Hager, H. Moriya, and N. Watanabe for discussions and comments on the manuscript. This study is supported by Grant-in-Aid for JSPS Overseas Research Fellow (20160228) and ERL, MIT.

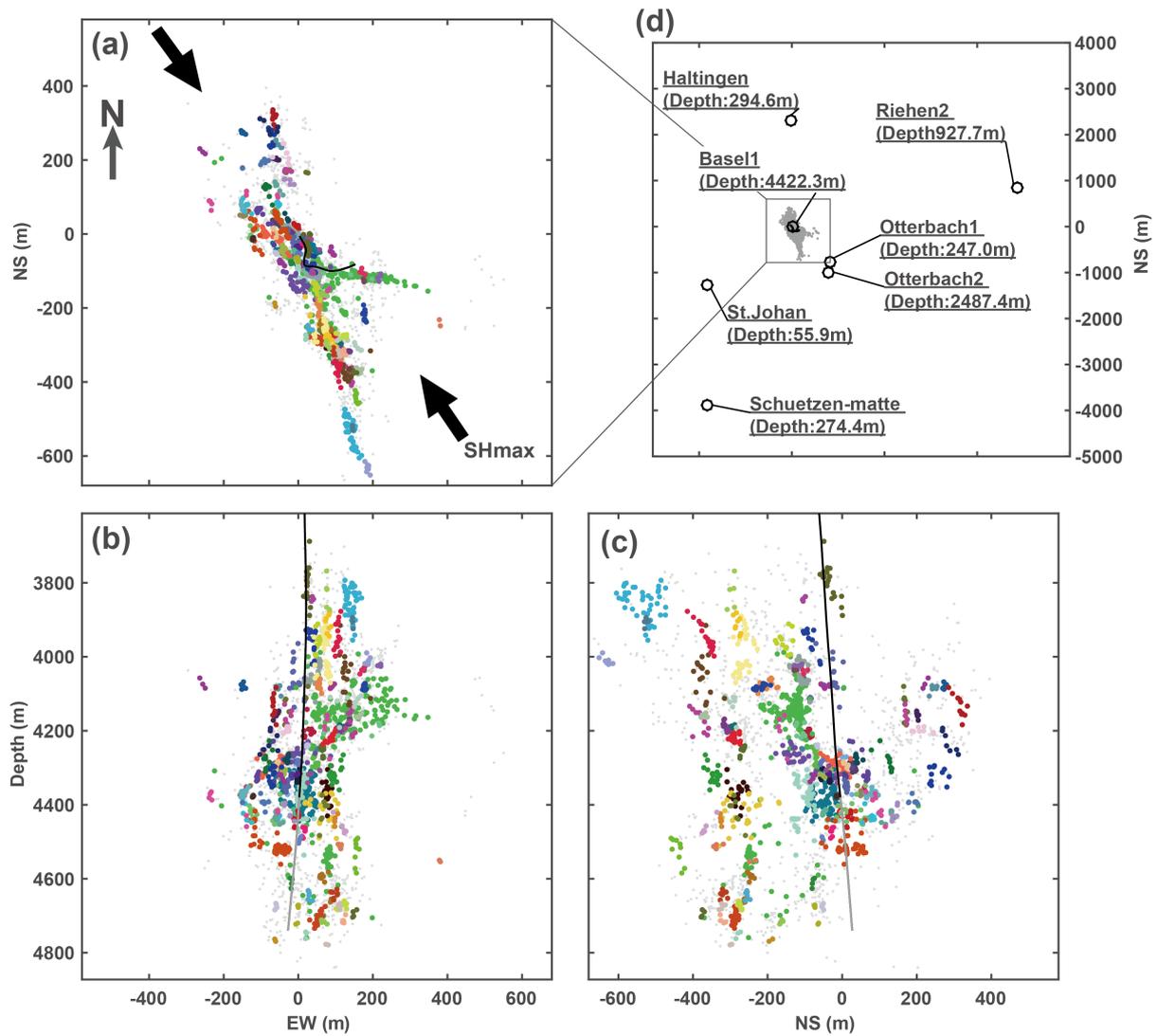

**Fig. 1.** Spatial distribution of microseismicity and seismic clusters based on waveform similarity. (a), plan view, (b), EW cross section (East is a positive direction), and (c), NS-cross section (North is in the positive direction). Events in each seismic cluster are marked with different colors. Events not belonging to a cluster are plotted with grey dots. Inset in (a): (orange) rose diagram showing the frequency distribution of strikes of seismic clusters, (purple) rose diagram showing the frequency distribution of strikes of each seismic event taken as a member of a seismic cluster. The black line is the trajectory of the injection well. The black part of the line is the cased section and the grey part is open hole section. (d), histograms showing the seismic cluster geometry characterization in terms of the ratios of eigenvalues associated with the cluster estimated from 3D principal component analysis on hypocenter distribution for each seismic cluster. Each histogram shows the counts of the contribution ratio of eigen values. A ratio of 1 for the largest eigenvalue would mean that the events in a cluster occur along a line. A cluster having two large eigenvalue ratios and a small value for the third ratio occurs along a plane. The orientation of the plane can be determined from the PCA analysis.

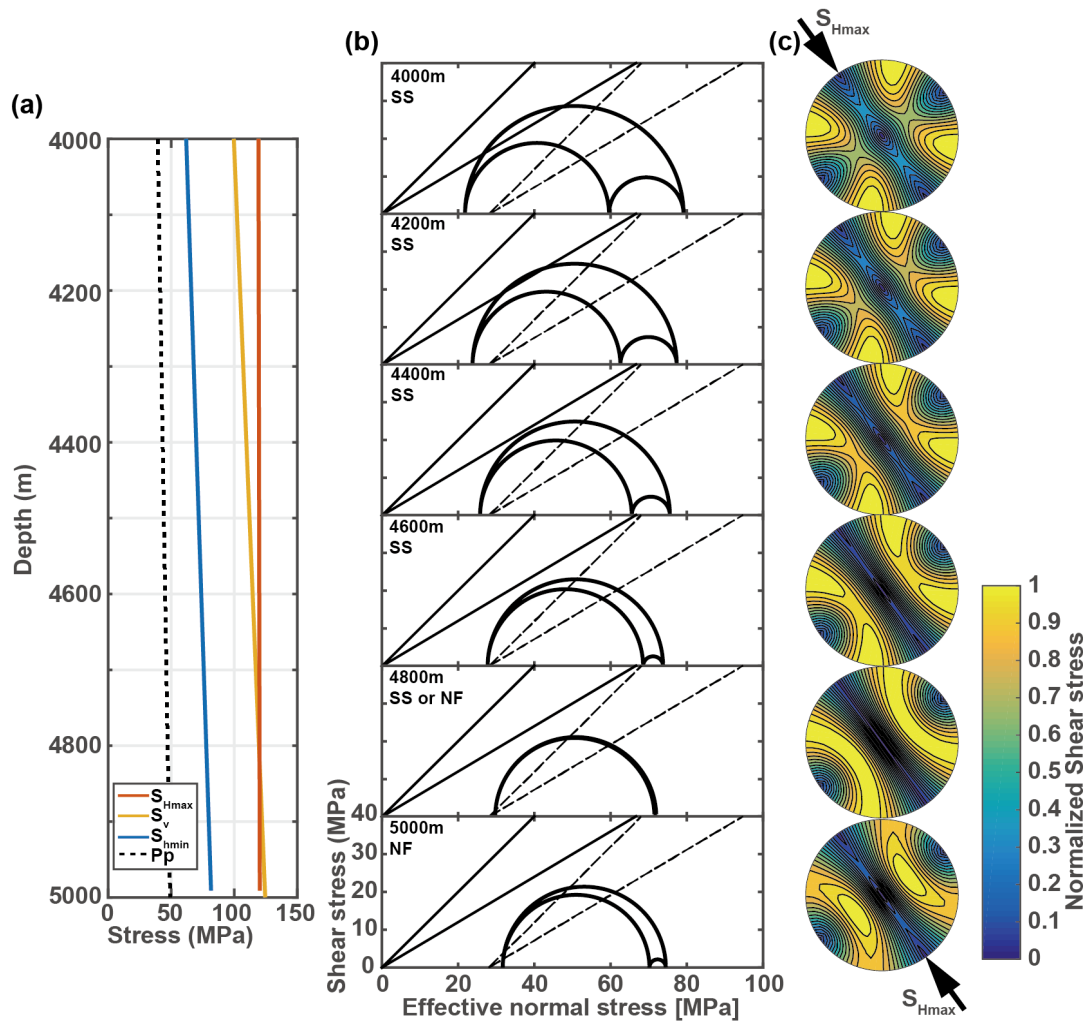

**Fig. 2.** Stress information used in this study. (a) Stress magnitude profile as a function of depth. (b) Mohr stress circles at selected depths with Coulomb failure criterion line for friction coefficient 0.6 and 1.0 (solid lines). Broken lines show failure criterion with pressure of 30 MPa. Stress state (SS: strike slip, NS: normal fault slip) is shown in each panel. (c) Lower hemisphere plots showing normalized shear stress for poles to planes at each depth. Vertical position of each panel of b and c corresponds the same depth as labeled in b.

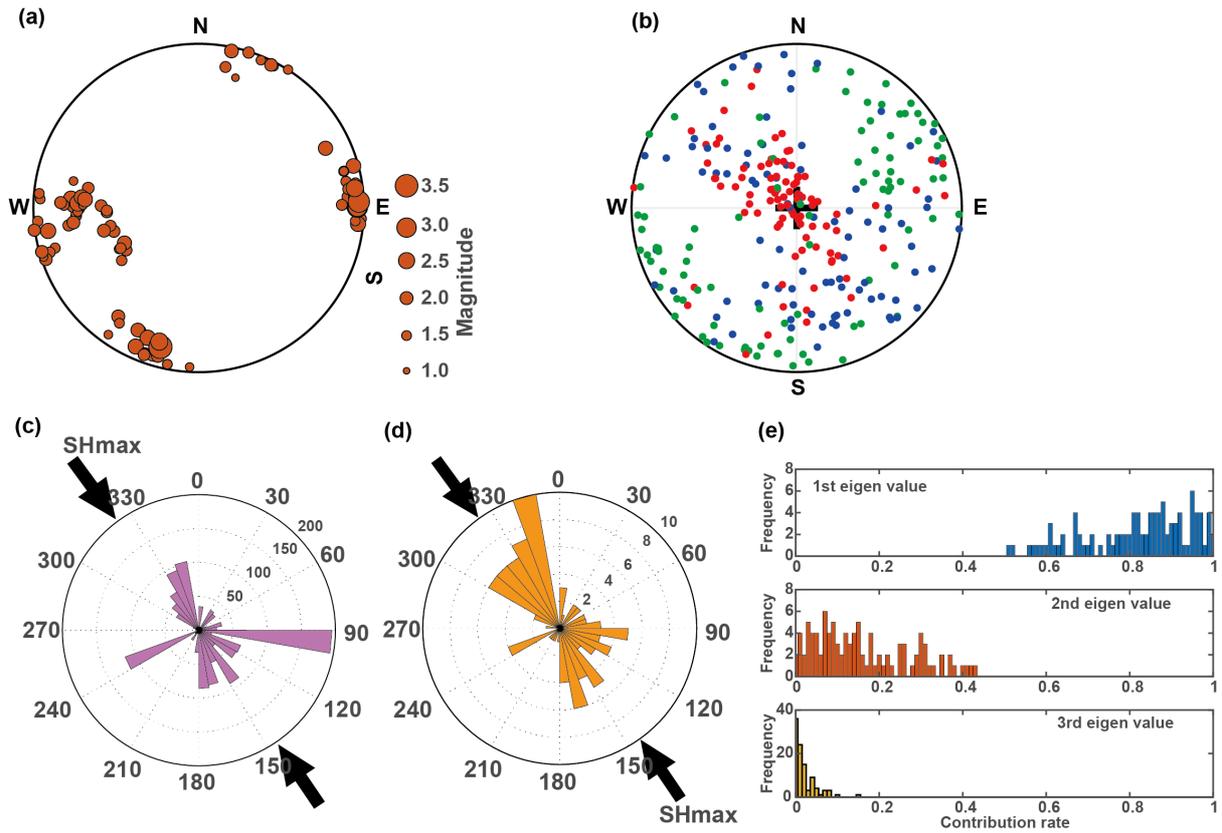

**Fig. 3.** Pole distribution of fault planes used in this study in lower hemisphere plot. (a) Pole distribution of selected slip plane from FPSs estimated by SED. The sizes of the circles scale with event magnitude. (b) Pole distribution for the planes identified from multiplet clusters consisting of more than five events (green dots). Red dots, and blue dots indicate the orientation of 1$^{st}$ and 2$^{nd}$ eigen vectors of multiplet clusters respectively.

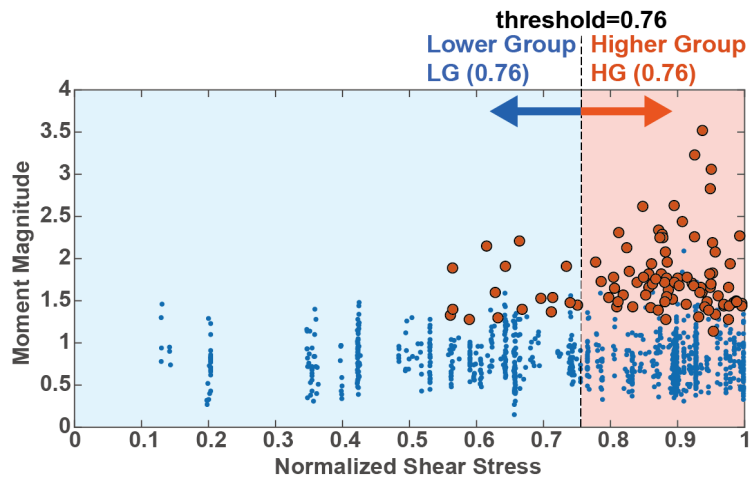

**Fig. 4.** Correlation between magnitude of induced seismicity and Normalized Shear Stress. Red dots correspond the events having FPS by SED and blue dots are the events whose fault geometry was estimated by cluster analysis. As an example, the case of threshold = 0.76 is shown. The events are subdivided into two groups; those having NSS higher than the threshold identified as HG (0.76) and those with NSS lower than the threshold identified as LG (0.76). Then, $b$-values for each group are estimated Mc estimated simultaneously. NSS threshold varies in subsequent analysis.

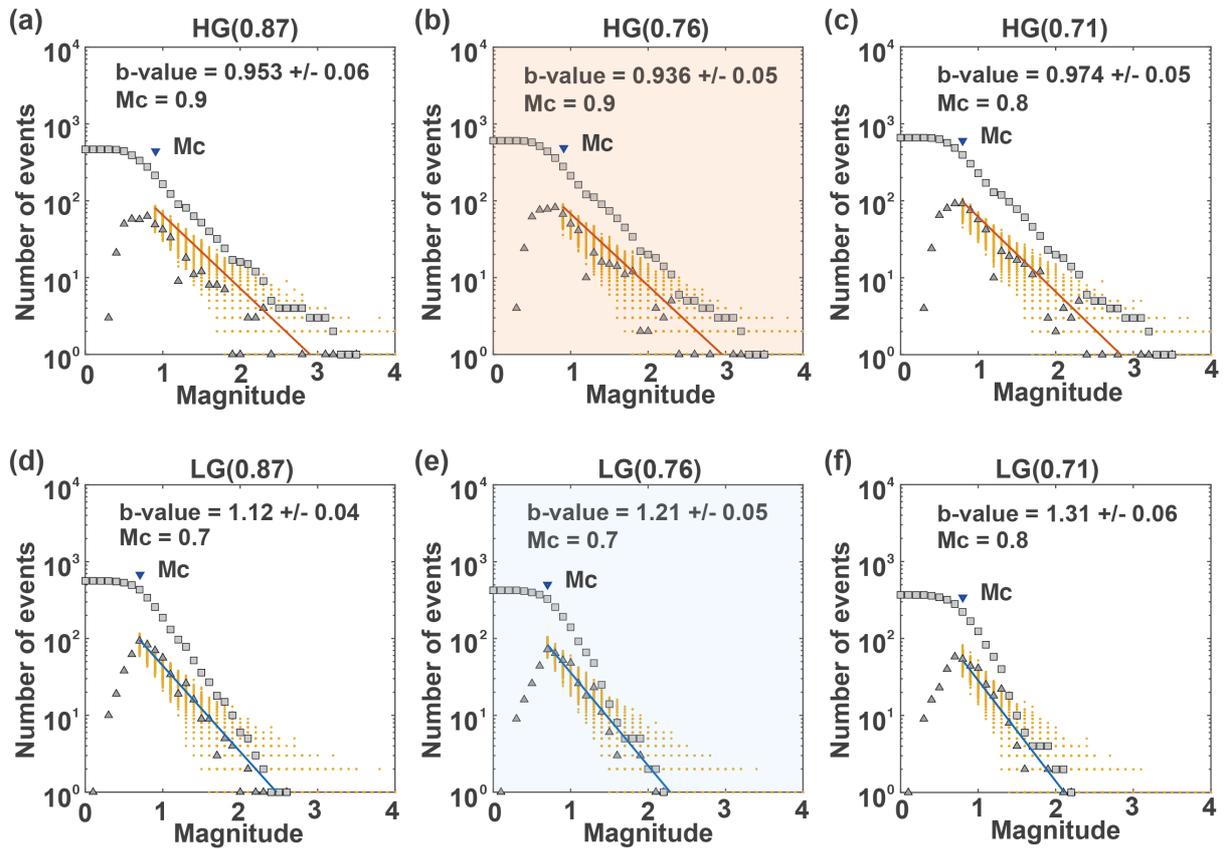

**Fig. 5**. Magnitude frequency distribution of induced seismicity occurring at various stress states. (a)-(c), NSS higher than given three values. (d)-(f), NSS lower than the criterion. Pairs with the same NSS thresholds are (b) and (e); (c) and (f); and (d) and (g). Blue triangles show the cutoff magnitude Mc for each plot. Squares show the cumulative number of events and triangles show the binned number of the events with 0.1 bin size. Mc and *b*-value for each catalog are shown in each panel. Yellow dots are the distribution of binned number of events from 1000 synthetic catalogs estimated using the estimated b-values.

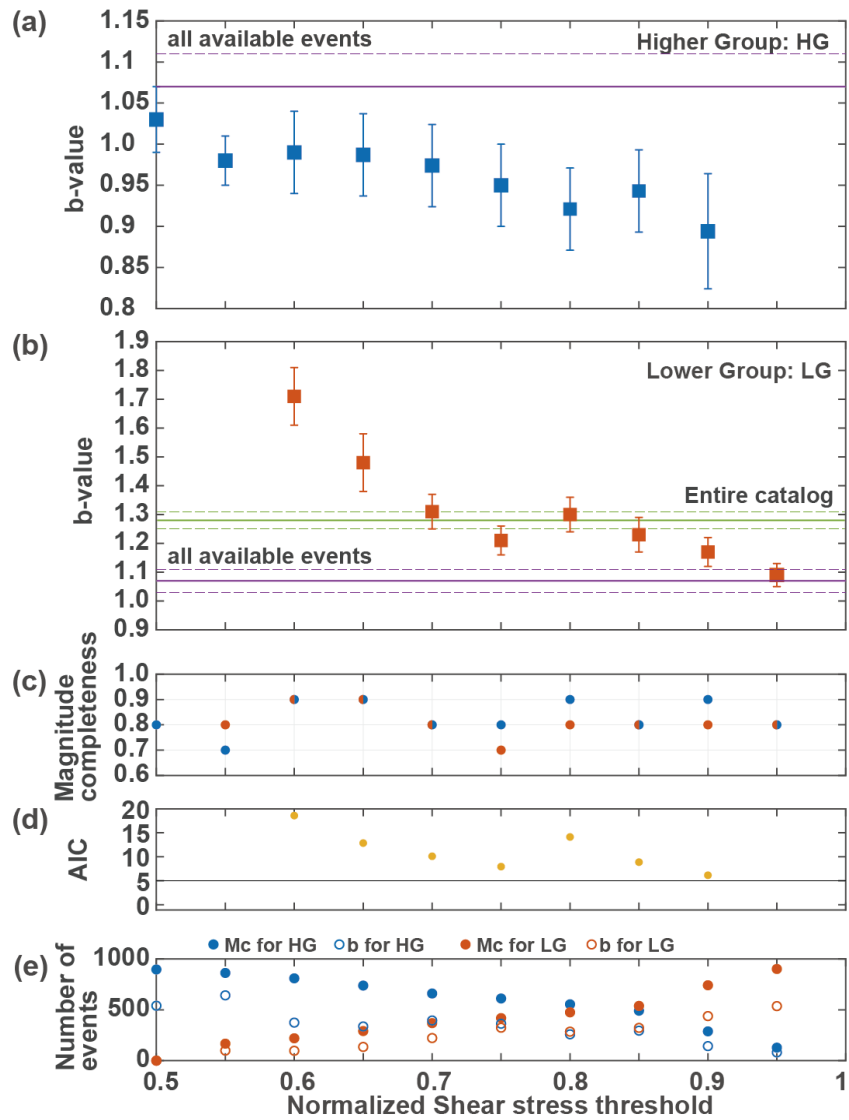

**Fig. 6.** *b*-value dependency on Normalized shear stress threshold. (**a**) *b*-values estimated for events in HG groups as a function of NSS. Blue squares with error bars show the *b*-values every 0.05 NSS. Purple and broken lines correspond to *b*-value and its uncertainty estimated from all events having known NSS. (**b**), *b*-value dependency on shear stress for events in LG groups. *b*-values for every 0.5 NSS are shown. Green and broken lines correspond to *b*-value and its uncertainty estimated from all events, and purple lines are same as in A. (**c**) Magnitude of completeness estimated for each sub-catalog with entire magnitude range method. Blue dots are hidden behind of red dots in some cases. (d) Significance of the difference between *b*-values of HG (blue squares in (a)) and LG (red squares in (b)) with same NSS threshold obtained using the Akaike information criterion (AIC). The highly

significant line of AIC=5 is shown with a black line. **(e)** Number of events used to estimate Mcs (filled circle) and *b*-values (open circle) for each HG (blue) and LG (red) group.

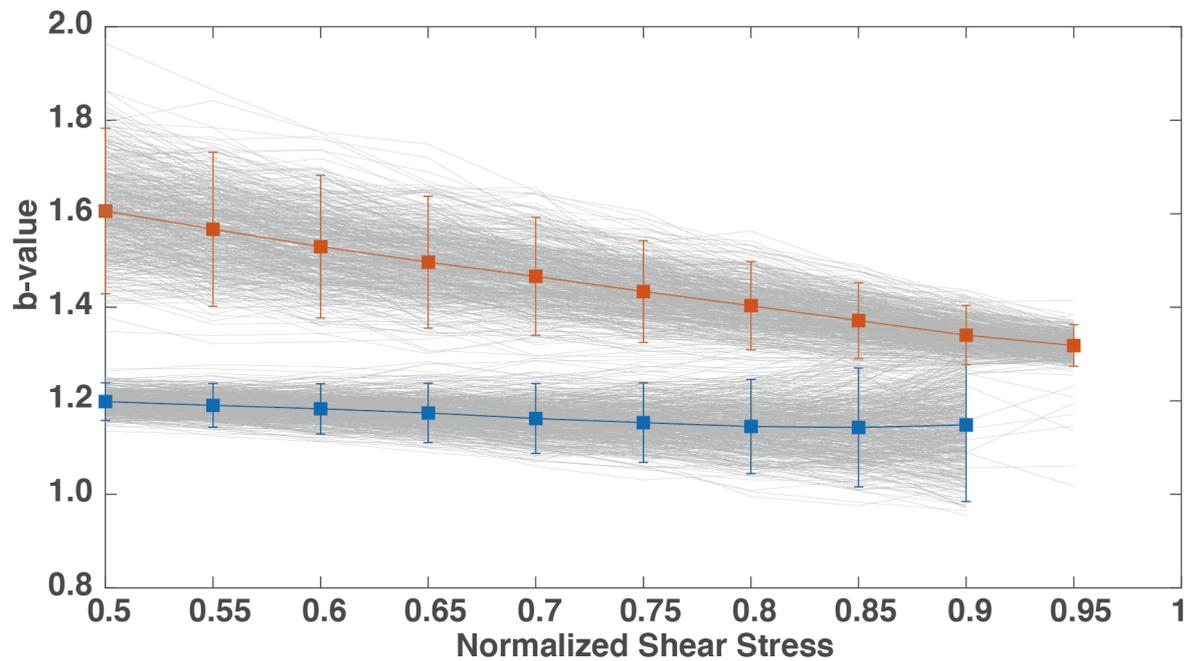

**Fig. 7.** *b*-value dependency on Normalized shear stress threshold under non-uniform stress state condition. Red and blue squares are the average of the 1000 b-value estimates for LG and HG, respectively. Grey lines are the results of b-value estimates from 1000 in-situ stress models considering the uncertainty of the stress model. 500 of 1000 results are shown. The error bars indicate the two sigmas (standard deviation) from the distribution of b-value estimates, not the uncertainty of b-value estimates.

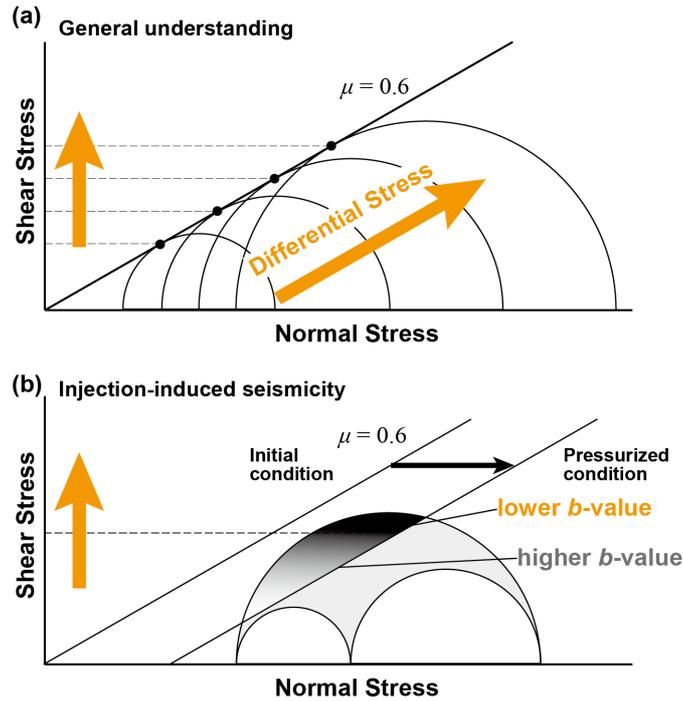

**Fig. 8.** Conceptual explanation of *b*-value dependency on shear stress. (a) Model for conventional case that *b*-value depends on differential stress. Increasing differential stress is expressed the increasing diameter of Mohr stress circle. Stress states of the well oriented planes for each differential stress is shown with black dots and their shear stress levels are projected to the vertical shear stress axis. Coulomb failure line shows the case of friction coefficient 0.6. Yellow arrows show conceptual correlation of *b*-value reduction. (b) injection induced seismicity case considered in this study. Differential stress is assumed as constant and proper intermediate stress is given. Shifted Coulomb failure line corresponds the pressurized condition. All existing fracture situated left of the shifted Coulomb failure line (grayscale gradation area) can have a shear slip. Broken line is the example of NSS threshold in case of NSS threshold = 0.87. Events in the critical region above the NSS threshold (black shaded area) show lower *b*-value and those below the NSS threshold (grey and white shaded area) show higher *b*-value. Yellow arrows also indicate the conceptual correlation of *b*-value dependency on shear stress.

# Supplementary information

# for

**Injection induced seismicity size distribution dependent on shear stress**

Yusuke Mukuhira*[1,2], Michael C. Fehler[1], Takatoshi Ito[2], Hiroshi Asanuma[3], Markus O. Häring[4]

[1]Earth Resources Laboratory, Department of Earth, Atmospheric and Planetary Sciences, Massachusetts Institute of Technology, Cambridge, Massachusetts, 02139, USA

[2]Institute of Fluid Science, Tohoku University, 2-1-1 Katahira, Aoba-ku, Sendai, 980-8577, Japan

[3]Fukushima Renewable Energy Institute, National Institute of Advanced Industrial Science and Technology (AIST), 2-2-9 Machiike-dai, Koriyama, Fukushima 963-0298, Japan.

[4]Häring Geo-Project, Wasserturmplatz 1, CH-4410, Liestal, Switzerland.

**Supplementary Materials**
Supplementary Figures 1 to
Supplementary Text

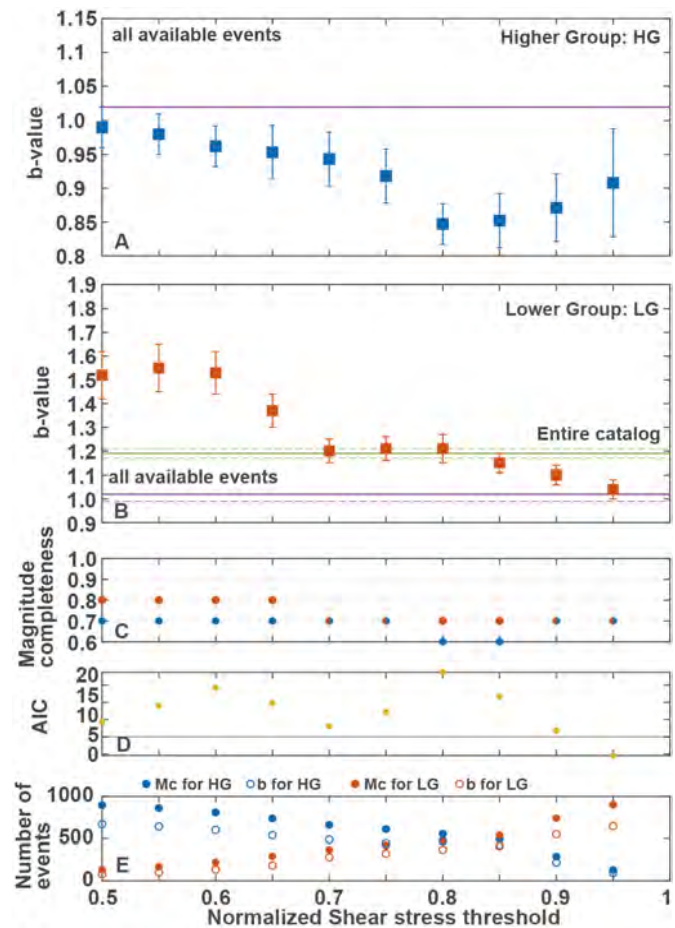

**Supplementary Fig. 1.** *b*-value dependency on Normalized shear stress threshold for Mc estimated using the maximum curvature method. Results of same analysis shown in Fig. 6 of the main text, but Mc was estimated using the maximum curvature method (Woessner and Wiemer, 2005), which often underestimates Mc to be around 0.2 lower than that estimated by other methods such as the EMR method (Woessner and Wiemer, 2005).

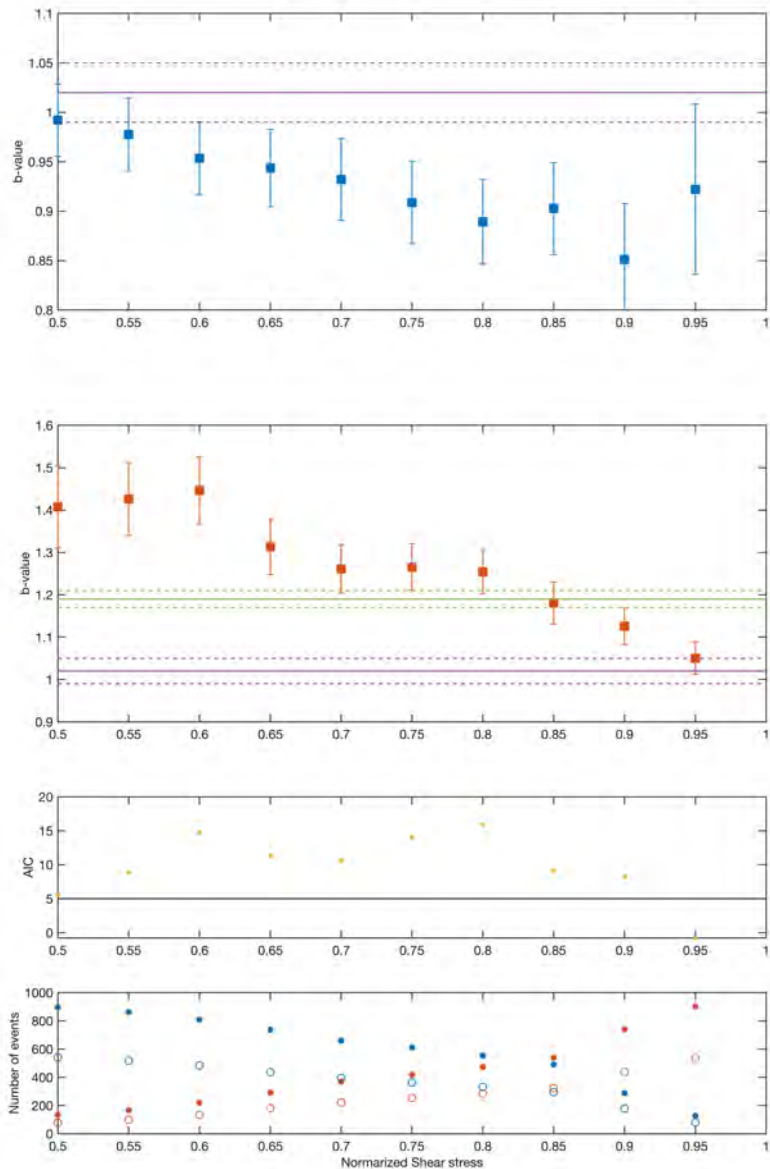

**Supplementary Fig. 2.** *b*-value dependency on Normalized shear stress threshold for constant Mc. Results of same analysis shown in Fig. 6 of the main text, but here, we used a constant Mc of 0.7, which is the Mc estimated with maximum curvature method for all available events. The uncertainties in *b*-value here are estimated with the method proposed by Shi and Bolt, (1982).

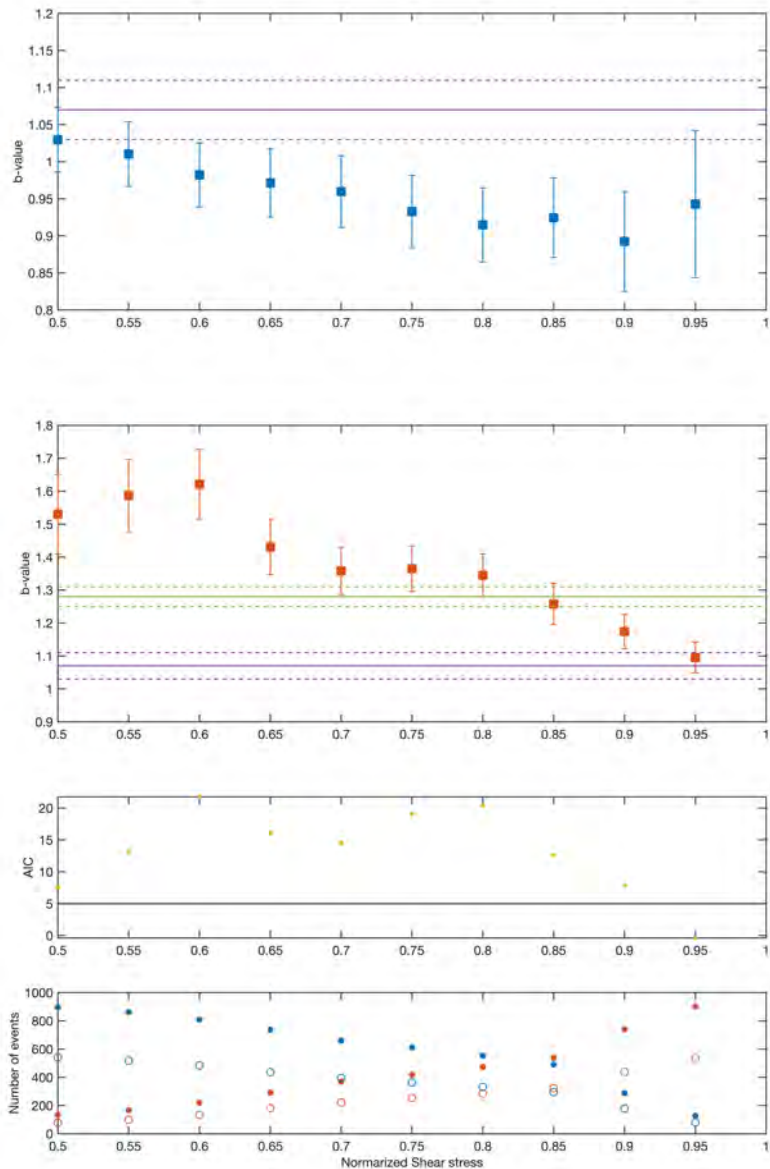

**Supplementary Fig. 3.** *b*-value dependency on Normalized shear stress threshold for constant Mc. Results of same analysis shown in Fig. 6 of the main text, but here, we used a constant Mc of 0.8, which is the Mc estimated with EMR method for all available events. The uncertainties in *b*-value here are estimated with the method proposed by Shi and Bolt, (1982).

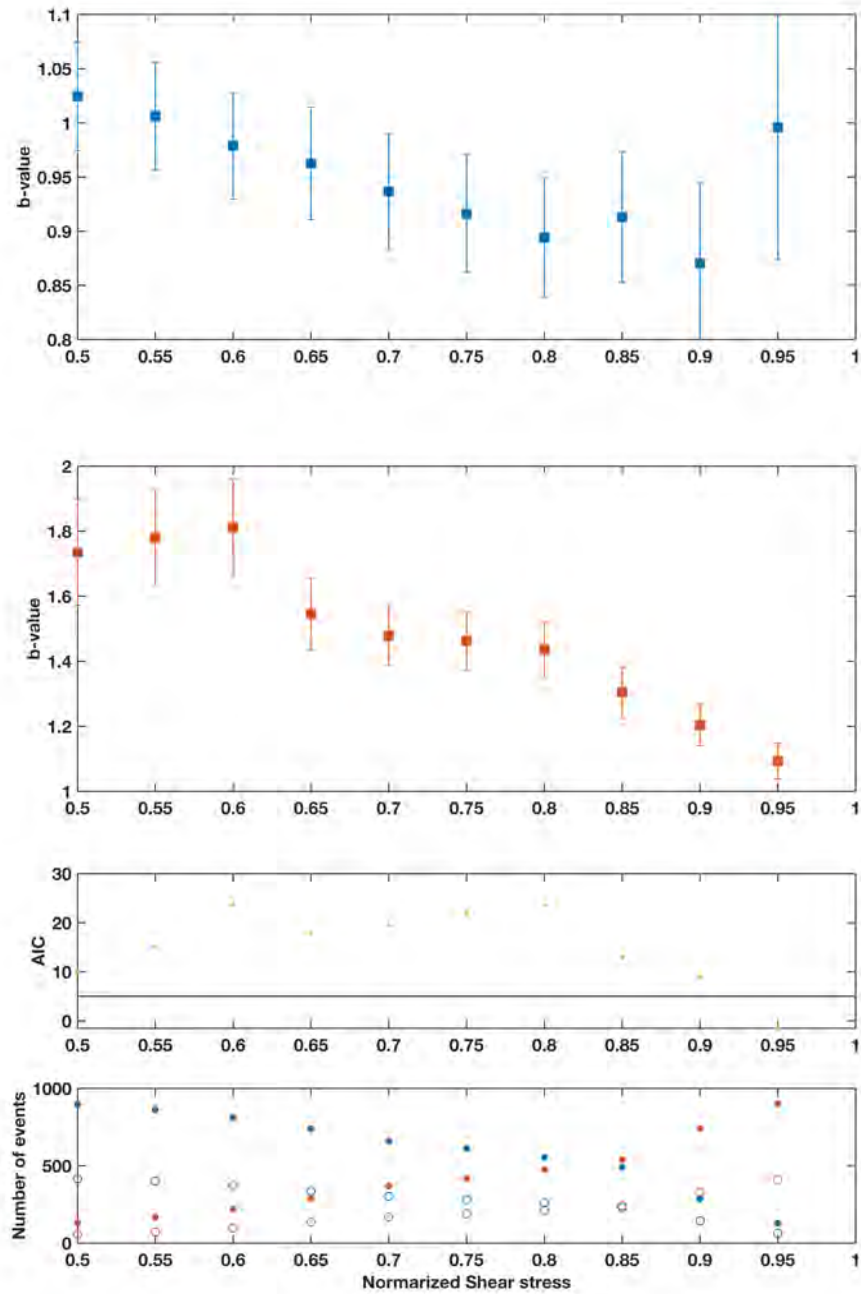

**Supplementary Fig. 4.** *b*-value dependency on Normalized shear stress threshold for constant Mc. Results of same analysis shown in Fig. 6 of the main text, but here, we used a constant Mc of 0.9, which is the Mc estimated with maximum curvature method for all available events. The uncertainties in *b*-value here are estimated with the method proposed by Shi and Bolt, (1982).

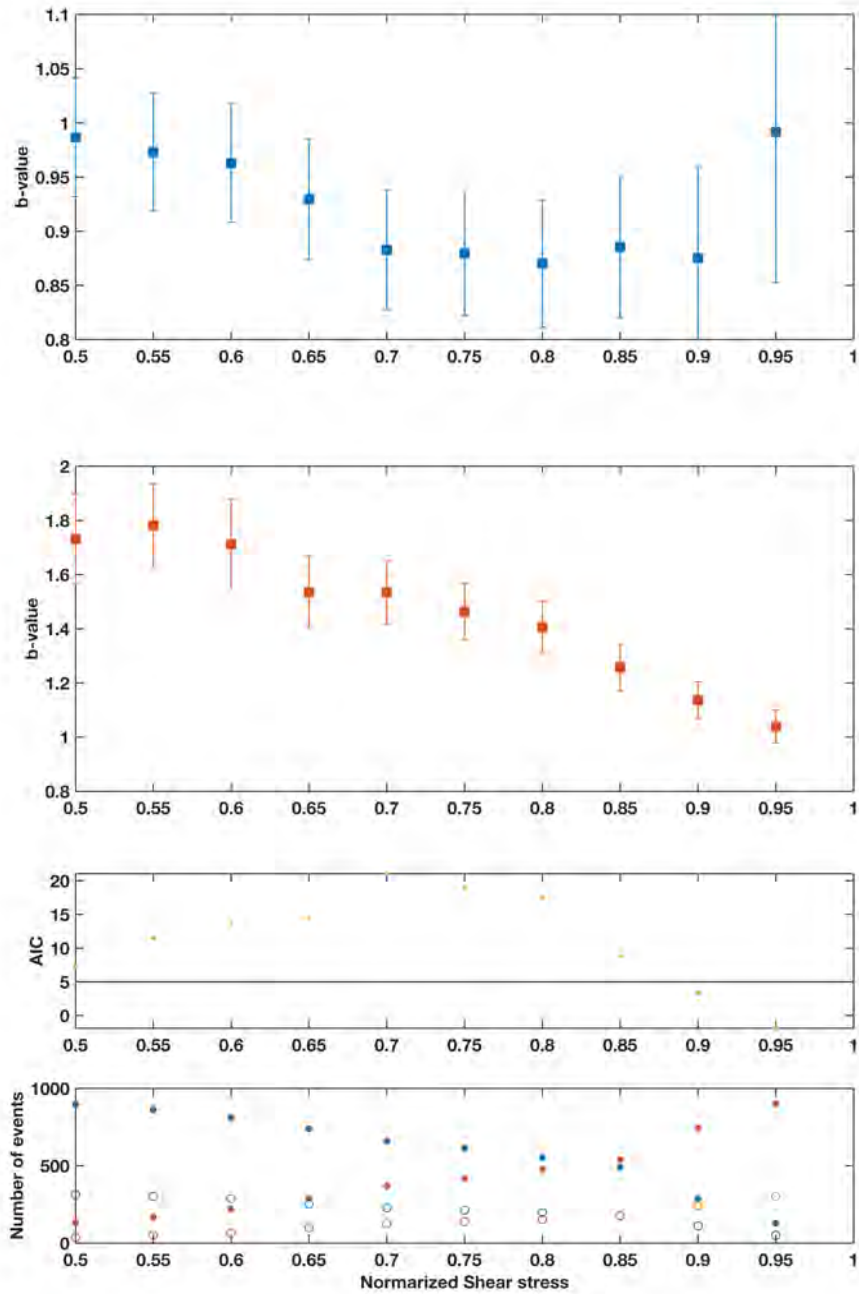

**Supplementary Fig. 5.** *b*-value dependency on Normalized shear stress threshold for constant Mc. Results of same analysis shown in Fig. 6 of the main text, but here, we used a constant Mc of 1.0, which is the Mc estimated with maximum curvature method for all available events. The uncertainties in *b*-value here are estimated with the method proposed by Shi and Bolt, (1982).

**Statistical test for the effect of the events without NSS information**

We know slip planes for a total of 1000 events from our analysis of focal mechanisms provided by SED and the multiplet analysis. Those events are used in our analysis of the relation between *b*-value and NSS. Here, we investigate the influence on our results of the remaining approximately 1400 events that do not have NSS information. Our investigation uses statistical sampling of those events.

**Procedure**

1. Extract NSS distribution from the events with known NSS information.
2. Randomly assign synthetic NSS values to events without NSS information according to the distribution of NSS from events with NSS information.
3. Merge the events with synthetic NSS and events with known NSS to make a semi-synthetic catalog.
4. Repeat 2 and 3, then make 1000 realizations of semi-synthetic catalogs.
5. Compute *b*-values for all of the 1000 synthetic catalogs while changing NSS threshold.

**Results**

- Most large events have NSS information; those without NSS information tend to have smaller magnitudes (from Supplementary Fig. 6).
- We computed the synthetic NSSs. An example of a distribution for the synthetic NSS values is shown by the purple histogram in Supplementary Fig. 7.
- For 1000 semi-synthetic catalogs, *b*-values for HG and LG are always separated suggesting that it is unlikely that the characteristics of HG and LG vary due to the effect of the events without NSS.

**Summary**

Supplementary Fig. 8 shows the distribution of *b*-values for HG and LG for various values of NSS cutoff. We still find three main observations that 1) *b*-value of HG decreases with increasing NSS threshold, 2) *b*-value of LG decreases with increasing NSS threshold, and 3) *b*-value of HG is always lower than b-value of LG. These observations are preserved even considering the effect of the events without NSS. Therefore, we conclude that our results are not impacted by the use of only events with known NSS value.

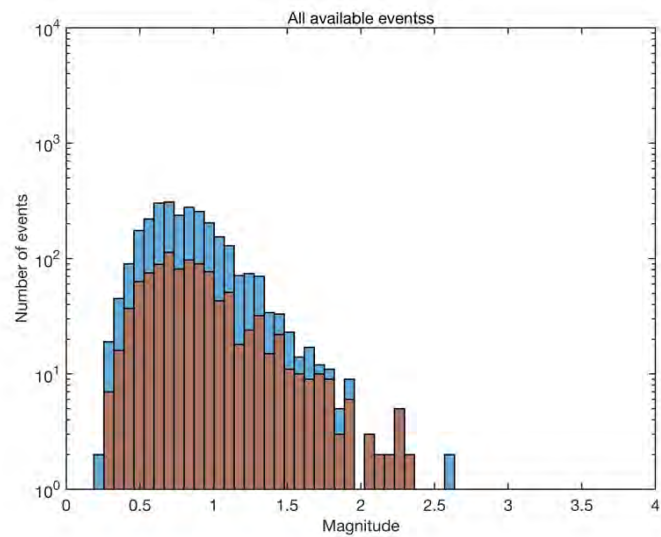

**Supplementary Fig. 6.** Magnitude frequency distribution of events with and without NSS information. Red histogram shows magnitude frequency distribution of events with NSS information and blue one shows magnitude frequency distribution of all events.

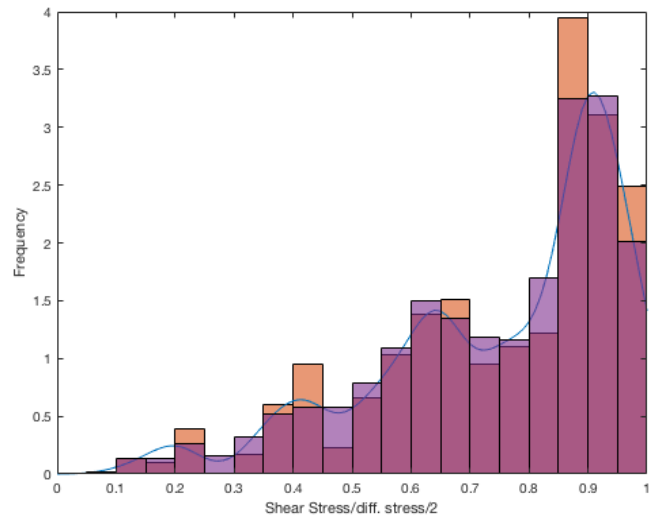

**Supplementary Fig. 7.** Histogram of observed NSS and forecasted NSS. Red histogram is the observed NSS. Blue line is probability distribution function of NSS. Purple histogram is just one example of synthetic NSS.

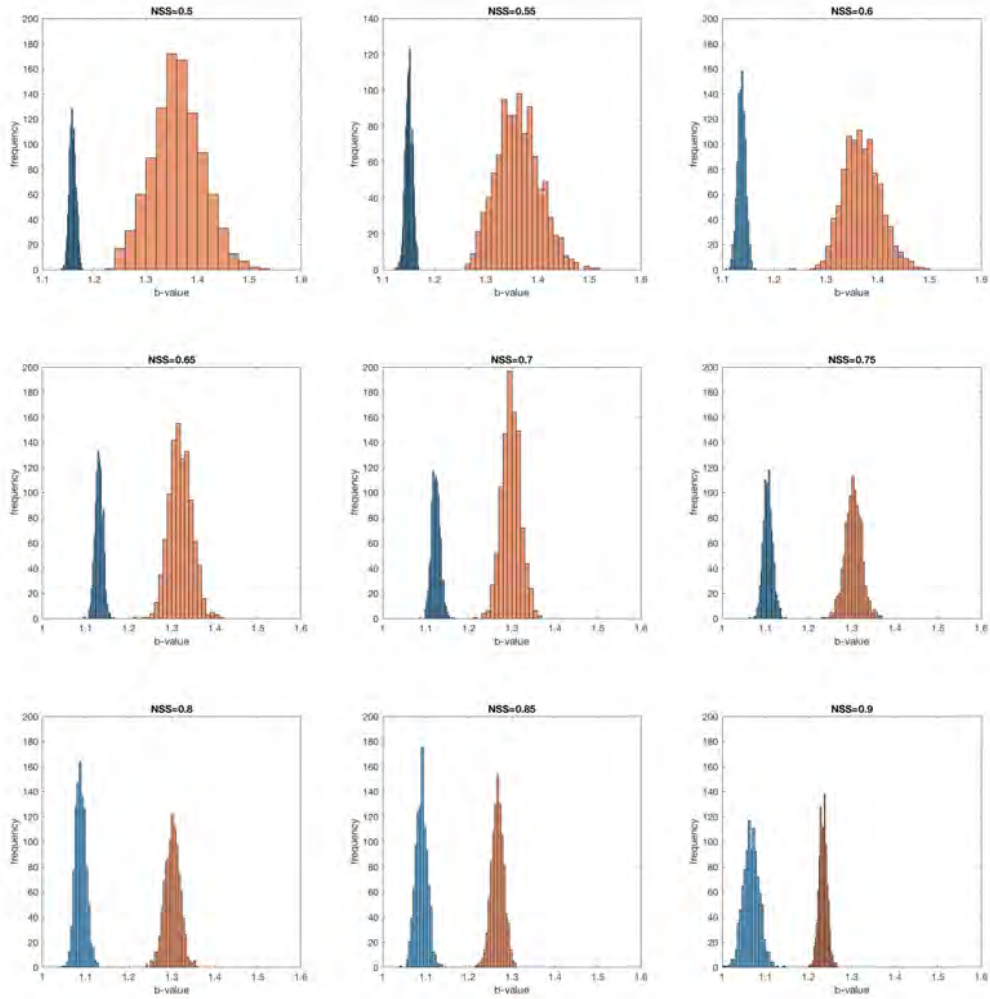

**Supplementary Fig. 8.** Results of statistical test. Distribution of observed *b*-values obtained by randomly assigning NSS values to events without NSS values using the NSS distribution of events with known NSS values. A total of 1000 random tests were evaluated. Blue shows distribution of *b*-value for LG events and red shows that for HG events. NSS threshold varies for each plot and is given at top of each plot.